\documentclass[twocolumn,showpacs,aps,prb]{revtex4}
\usepackage{graphicx}
\usepackage{dcolumn}
\usepackage{bm}
\usepackage{epsf}

\newcommand{\edc}{\end{document}}
\newcommand{\bb} {\begin{thebibliography}}
\newcommand{\eb}{\end{thebibliography}}
\newcommand{\bi}[1]{\bibitem{#1}}
\newcommand{\bc}{\begin{center}}
\newcommand{\ec}{\end{center}}

\newcommand{\be}{\begin{equation}\small}
\newcommand{\ee}{\end{equation}\normalsize}
\newcommand{\bea}{\begin{eqnarray}}
\newcommand{\eea}{\end{eqnarray}}
\newcommand{\ba}{\begin{array}{l}   }

\newcommand{\ea}{\end{array}}

\begin{document}
\title{Comments on ``Thermally induced rotons in two-dimensional dilute Bose gases"}
 \author{Sang-Hoon  \surname{Kim}} \email{shkim@mmu.ac.kr }
\affiliation{Division of Liberal Arts $\&$ Sciences,
Mokpo National Maritime University,
Mokpo 530-729, Republic of Korea}
\maketitle

In a study of interacting Bose particles
the effective interaction, $g$,
which  contains pairwise interaction between atoms
is an essential factor.
In 3D it depends on the scattering length  only such as
$g(3)=4 \pi \hbar^2 a/m$,
where $m$ is the atomic mass and
$a$ is the s-wave scattering length.
However, in general, it depends on momentum, too.
In $D<4$ dimensional case, the form of the
D-dimensional effective interaction $g(D)$
 has been suggested by  Nogueira and Kleinert in 2006.\cite{nogueira}
The final form is shown at the Eq. (8) of the paper as
 \be
 g(D)=\frac{4 \pi^{D/2} \hbar^2 a^{D-2}}{m}\frac{1}{2^{2-D}\Gamma(1-\frac{D}{2}) \gamma^{D-2} + \Gamma(\frac{D}{2}-1)},
\label{1}
 \ee
where $\Gamma$ is the Gamma function and
 $\gamma$ is the dimensionless  gas parameter in D-dimension
 given by $\gamma=n a^D \ll 1$,
 where $n$ is the D-dimensional gas density.

However, actually, the above form of $g(D)$ is not correct.
The correct form is slightly different from Eq. (\ref{1})
in the exponent of the $\gamma$. It is
\be
 g(D)=\frac{4 \pi^{D/2} \hbar^2 a^{D-2}}{m}\frac{1}{2^{2-D}\Gamma(1-\frac{D}{2}) \gamma^{D/2-1} + \Gamma(\frac{D}{2}-1)}.
\label{3}
 \ee
 The mistake of the Eq. (\ref{1}) came from the error
 that in the final step of the derivation of the paper
 they chose $pa \rightarrow na^D = \gamma $
instead of $pa \rightarrow \sqrt{na^D} = \sqrt{\gamma}$,
where $p$ is the momentum.

The mistake is trivial but a direct substitution of the formula in Eq. (\ref{1}) into any D-dimensional problem leads a wrong result.
For example, the Eq. (\ref{1}) cannot reproduce
the well-known 2D results of Bosons by M. Schick.\cite{schick}
Only the formula in Eq. (\ref{3}) in a D-dimensional Boson research
produces correct ones.\cite{kim2}
Average theoretical physicist who uses the formula $g(D)$
would not make the mistake,
but we feel to comment for that because its role of $g(D)$ is critical
 in D-dimensional researches.

\bb{99}
\bi{nogueira} F. S. Nogueira and H. Kleinert, \prb {\bf 73}, 104515 (2006).
\bi{schick} M. Schick, \pra {\bf 3}, 1067 (1971).
\bi{kim2} S.-H. Kim and M. P. Das, arXiv:0808.3288v1 [cond-mat.stat-mech] (2008).
\eb
\end{document}